\documentclass[aps,prl,twocolumn,showpacs,preprintnumbers,amsmath,superscriptaddress,amssymb,floats,nofootinbib]{revtex4}

\usepackage{graphicx}
\usepackage{dcolumn}
\begin{document}

\def\Journal#1#2#3#4{{#1} {\bf #2}, #3 (#4)}
\def\NCA{\rm Nuovo Cimento}
\def\NIM{\rm Nucl. Instrum. Methods}
\def\NIMA{{\rm Nucl. Instrum. Methods} A}
\def\NPB{{\rm Nucl. Phys.} B}
\def\PLB{{\rm Phys. Lett.}  B}
\def\PRL{\rm Phys. Rev. Lett.}
\def\PRD{{\rm Phys. Rev.} D}
\def\PRC{{\rm Phys. Rev.} C}
\def\ZPC{{\rm Z. Phys.} C}
\def\JPG{{\rm J. Phys.} G}
\def\st{\scriptstyle}
\def\sst{\scriptscriptstyle}
\def\mco{\multicolumn}
\def\epp{\epsilon^{\prime}}
\def\vep{\varepsilon}
\def\ra{\rightarrow}
\def\ppg{\pi^+\pi^-\gamma}
\def\vp{{\bf p}}
\def\ko{K^0}
\def\kb{\bar{K^0}}
\def\al{\alpha}
\def\ab{\bar{\alpha}}
\def\be{\begin{equation}}
\def\ee{\end{equation}}
\def\bea{\begin{eqnarray}}
\def\eea{\end{eqnarray}}
\def\CPbar{\hbox{{\rm CP}\hskip-1.80em{/}}}
\title{\large \bf Relaxation of Spin Polarized $^3$He in Mixtures of $^3$He and $^4$He\\ Below the $^4$He Lambda Point}
\author{Q.~Ye, D.~Dutta, H.~Gao, K.~Kramer, X.~Qian, X.~Zong\\
{\it Triangle Universities Nuclear Laboratory and \\
Department of Physics, Duke University, Durham, NC 27708,~USA}\\
L.~Hannelius,  R.~D.~McKeown, B.~Heyburn, S.~Singer\\
{\it Kellogg Radiation Laboratory and \\
Department of Physics, California Institute of Technology, Pasadena, CA 91125,~USA}\\
R.~Golub\\
{\it Department of Physics, North Carolina State University, Raleigh, NC 27695,~USA}\\
 E.~Korobkina\\
{\it Department of Nuclear Engineering, North Carolina State University, 
Raleigh, NC 27695,~USA}}

\begin{abstract}

We report a first study of the depolarization behavior of spin polarized
$^3$He in a mixture of $^3$He-$^4$He at a temperature below the $^4$He 
$\lambda$ point in a deuterated TetraPhenyl Butadiene-doped deuterated PolyStyrene (dTPB-dPS) coated acrylic cell. 
In our experiment the measured $^3$He relaxation time is due to the convolution of the $^3$He longitudinal relaxation time, $T_1$, and the diffusion time constant of $^3$He in superfluid $^4$He since depolarization takes place on the walls.
We have obtained a $^3$He relaxation time of $\sim$ 3000 seconds at a temperature around 1.9 K. We have shown that it's possible to achieve values of wall depolarization probability on the order of $(1 - 2)\times 10^{-7}$ for polarized $^3$He in the superfluid $^4$He from a dTPB-dPS coated acrylic surface.

\end{abstract}


\maketitle
Hyper-polarized $^3$He based on the technique of optical 
pumping~\cite{bou60,techs}
has found applications in diverse fields such as in the study of quantum 
phenomena in low temperature fluids~\cite{cmexp} and in the search for 
violations of fundamental symmetries~\cite{npdg,edm,bear00}. They are also 
routinely used as polarized neutron spin-filters~\cite{nsf}, as effective 
polarized neutron targets for nuclear and particle physics 
experiments~\cite{wxu00} and for low magnetic field magnetic 
resonance imaging~\cite{mri}. All such applications have motivated, as 
well as benefited from, studies of the relaxation mechanisms of 
polarized $^3$He in gas, liquid and superfluid phases and under different
surface conditions. The study of relaxation of 
polarized $^3$He in $^3$He - $^4$He mixtures at low temperatures is, 
however, of 
longstanding interest in its own right~\cite{he3lit}. The simple atomic 
structure of $^3$He makes a system of $^3$He atoms ideal to model
correlated fermions.

Super-thermal production of Ultra-Cold Neutrons (UCN)
from superfluid $^4$He has been 
demonstrated~\cite{ucn} as an efficient way of producing a large number of UCNs.
The capability of storing a large number of UCNs following their production 
is important for experiments studying fundamental properties of the neutron,
for example the experiment on the search of the
neutron electric dipole moment~\cite{edm}.
In this experiment, the
deuterated TetraPhenyl Butadiene-doped deuterated PolyStyrene 
(dTPB-dPS) coated acrylic 
surface is chosen for such an application 
because of the small neutron absorption rate on the surface
and its wavelength shifting property.
The focus of this work is a first study of polarized $^3$He relaxation
time in a mixture of $^3$He-$^4$He below the $^4$He
$\lambda$ point in a dTPB-dPS coated acrylic cell. Such a study may find 
applications in the development of cryogenic $^3$He magnetometers 
for experiments
where trapping of polarized UCNs is involved as well as
in other types of applications
where polarized $^3$He atoms are employed at low temperatures.
At present the feasibility of $^3$He magnetometers for UCNs has 
been studied only
at room temperature~\cite{magnetometer}.

While a number of experiments~\cite{mann,lowe,piegay,lusher,himbert} 
have reported $^3$He longitudinal relaxation times ($T_1$)
in mixtures of $^3$He-$^4$He at temperatures similar to our work, the 
measurements most relevant to ours are ~\cite{lusher,himbert,lowe}. 
It was observed that $^3$He atoms  
in a gaseous phase~\cite{lusher} in the presence of $^4$He had a longer
$T_1$ below temperatures where superfluid $^4$He film was formed.  
Lowe {\it et al.}~\cite{lowe} observed little $^3$He concentration 
dependence in the 
observed $^3$He $T_1$, which was shorter 
than 300 seconds in $^3$He-$^4$He solutions between 1.5 and 3.3 K.
In this paper we 
report the first results of the $^3$He relaxation time~\footnote{The measured $^3$He relaxation time is due to the convolution of the $^3$He longitudinal relaxation time, $T_1$, and the diffusion time constant of $^3$He in superfluid $^4$He.}  
in the presence of superfluid 
$^4$He film and liquid in a 
dTPB-dPS coated acrylic cell at a temperature of 1.9~K and at a magnetic 
field of 21 Gauss. 

\begin{figure}[htbp]
\includegraphics*[width=7.5cm]{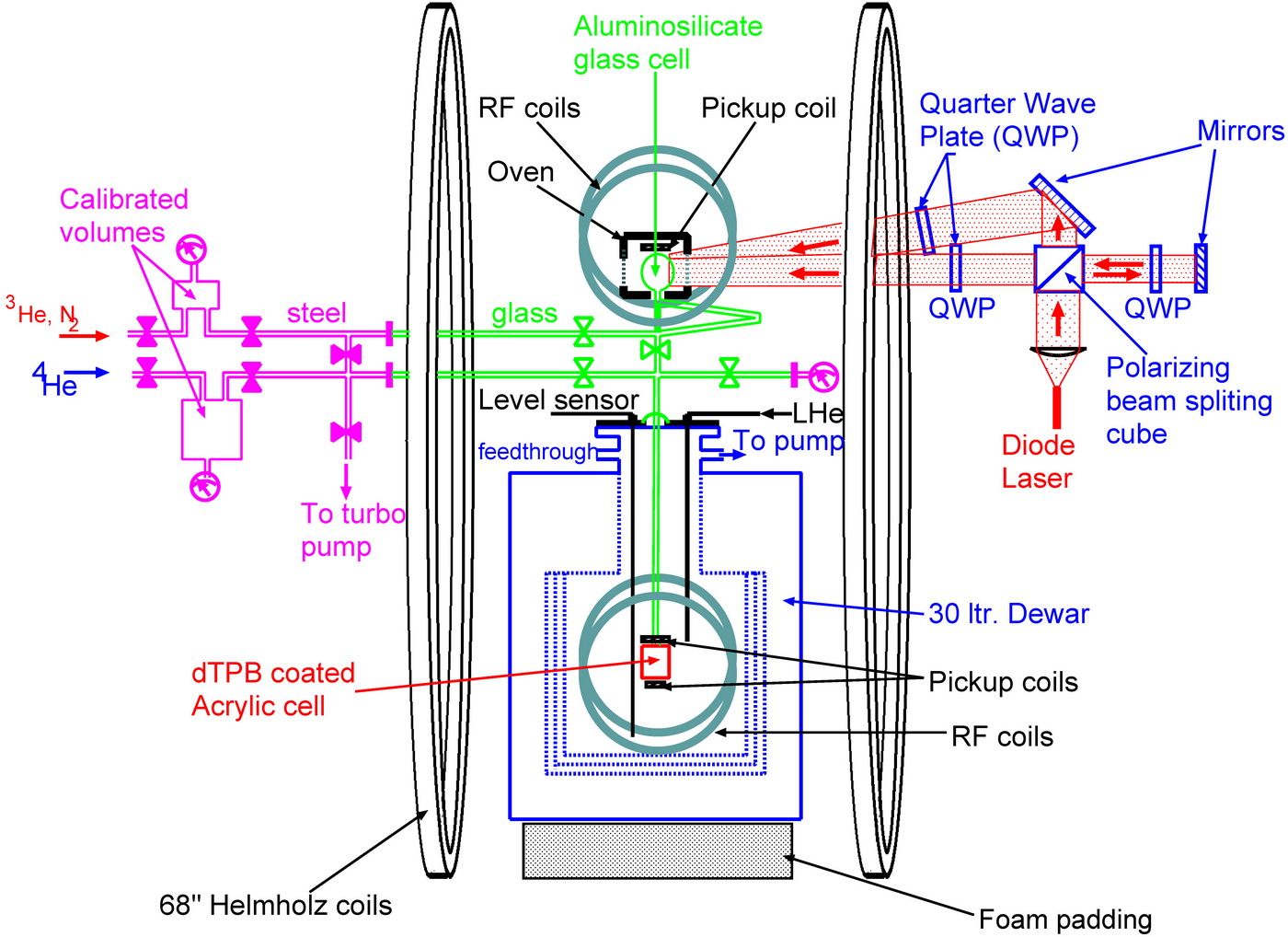}
\caption{A schematic of the experimental setup.} 
\label{apparatus}
\end{figure}

We have adopted the spin-exchange optical pumping (SEOP) technique for 
producing polarized $^3$He nuclei. The polarization is measured using the 
adiabatic fast passage (AFP) technique of Nuclear Magnetic Resonance (NMR). 
The schematic of the entire apparatus is shown in Figure~\ref{apparatus}. It 
consists of a pair of Helmholtz coils with a diameter of 68"
and the typical magnetic holding field is 21 Gauss.
A two-chamber apparatus for polarizing $^3$He nuclei and for 
measuring their relaxation time at 
cryogenic temperatures is constructed from aluminosilicate glass (GE180), 
and a cylindrical acrylic cell.
The two chambers are connected via a 3~mm diameter, 
21" long pyrex
capillary tubing and are separated by a glass valve. The top chamber is a 
spherical cell with a diameter of 2.0", while the bottom chamber 
is a cylindrical acrylic cell with an outer (inner) diameter of
2.0" (1.45") and a length of 2.0" attached to the glass via a 
0.5" long glass to copper seal with a diameter of 3 mm. The copper seal is 
attached to the acrylic cylinder using the low temperature epoxy 
Emerson \& Cuming Stycast 1266. The inner surface of the acrylic cell is 
coated with dTPB-dPS. Details of the coating procedure can be 
found in~\cite{coating}. Each chamber 
can be independently evacuated and filled with either $^3$He and nitrogen 
gas~\footnote{N$_2$ gas, introduced for efficient optical pumping, will freeze on the wall of the capillary tube at low temperatures.}
(top chamber) or $^4$He gas (bottom chamber), 
and they can be isolated 
from the gas handling system via a pair of glass valves. 
The temperature of 
the bottom cell can be lowered to $\sim$ 1.8~K by pumping on the 
vapor above the liquid helium inside the dewar. The temperature is 
measured using a calibrated cernox resistance thermometer. 

Two NMR systems are built in order to measure the $^3$He 
relaxation time in both the top (pumping) cell and the bottom cell. 
Each NMR setup consists of a pair of RF coils, 12" in diameter and one or two
pickup coils~\footnote{Two pickup coils are attached to the dTPB-dPS 
coated acrylic cell as shown in Figure~\ref{apparatus}.} with a diameter of 1.5". 
The RF and pickup coils for 
the low temperature (bottom) cell are placed inside the dewar and hence 
are immersed in liquid helium during the measurement cycle.


The top cell is prepared by baking it under vacuum at $\sim$ 350$^{\circ}$ C, and then distilling
 a few milligrams of Rb into the cell (also under vacuum). 
Once the top cell is ready, a known amount of $^3$He is introduced into the 
cell for each measurement (the amount can be varied as desired) and 
N$_2$ ($\sim$ 50 - 100 torr filled at room temperature) 
is also added as a buffer gas. 
The top cell is enclosed in an oven and heated to 190 degrees Celsius and 
a 30W circularly polarized laser light at 794.7~nm is incident onto the cell 
to polarize the $^3$He atoms through SEOP. While the $^3$He atoms in the top 
cell are being polarized, liquid $^4$He is filled into the dewar and the 
temperature of the bottom (acrylic) cell is lowered below the liquid $^4$He boiling 
temperature
by pumping on the $^4$He vapor with a large throughput pump. 
Once the acrylic cell has 
reached the desired temperature with the lowest temperature being 1.8 K, a 
known amount of $^4$He gas is introduced into the acrylic cell. 
The laser is then turned off and the top cell is cooled to
room temperature, after which the glass valve separating the two 
chambers is opened to allow the polarized $^3$He atoms to diffuse to the 
bottom acrylic cell. The N$_2$ gas condenses on the way down and does not
enter the bottom cell.
The valve is closed after 30 seconds and a series of NMR-AFP measurements are 
performed with a time interval between 50 to 220 seconds. 
The amount of $^3$He
in the capillary tube in our experiment is negligible.

\begin{figure}[htbp]
\includegraphics*[width=7.5cm]{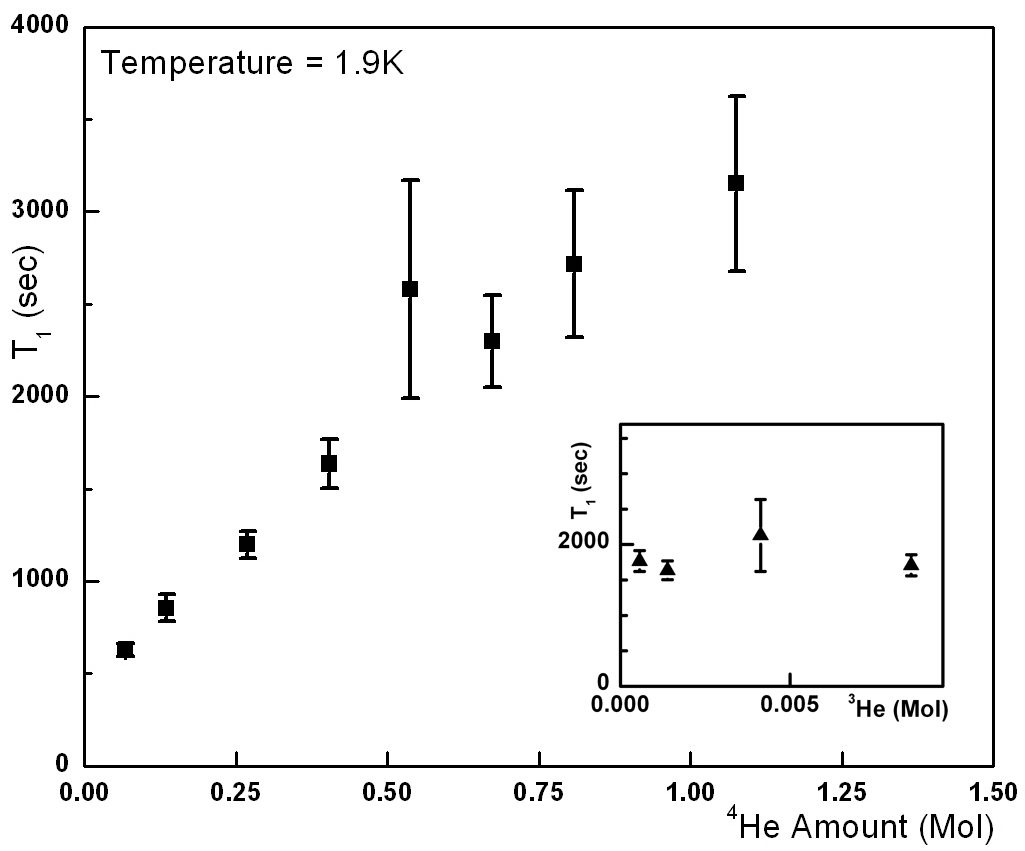}
\caption{The relaxation time of polarized $^3$He as a function of the amount of $^4$He in the measurement cell at a temperature of $<$~1.9~K$>$.
The error bars are the quadrature sum of the statistical and systematic uncertainties. The figure inset shows the polarized $^3$He relaxation time as a function of the amount of $^3$He, when the amount of  $^4$He is held constant at 0.404 mol.
}
\label{results}
\end{figure}

Measurements are carried out with a dTPB-dPS coated acrylic cell.
The relaxation time of $^3$He is 
consistently shorter than 10 seconds with no $^4$He inside the cell at 
a temperature of around 1.9~K.  
A strong correlation between the $^3$He relaxation time and 
the amount of $^4$He atoms introduced into the cell is observed.
Further, relaxation times in excess of 3000 seconds are observed. 
For comparison, the $^3$He relaxation times at room temperature from 
the optical pumping GE180 glass cell are between 5980 and 6700 seconds.
Figure~\ref{results} shows the $^3$He relaxation times at 21 gauss
holding field from a dTPB-dPS coated acrylic cell at $\sim$1.9 K. 
The amount of $^4$He is varied from 0.0 to 1.076 mol while the 
amount of $^3$He is fixed at 0.0014 mol\footnote{The temperature range for these measurements is from 1.83 to 1.90 K.}.
$^3$He relaxation times are extracted
by fitting the NMR data as a 
function of time to an exponential decay form with corrections for AFP spin 
flipping inefficiency. 
The AFP spin flipping inefficiency is determined to be $(1 \pm 1) \%$.
The longest $^3$He relaxation time obtained at $\sim$ 1.9 K 
from the dTPB-dPS coated acrylic cell is 3152 $\pm$ 86 (statistical) $\pm$ 473 
(systematic) seconds. The main 
contributions to $^3$He 
depolarization are the dipole-dipole relaxation mechanism, 
the magnetic field gradient 
effect and the surface effect at the wall.
For the data shown in Figure~\ref{results} the dipole-dipole relaxation time is calculated to be $6.24 \times 10^{5}$ seconds~\cite{newbury}.
The relaxation time due to the magnetic field gradient in our system is 
studied carefully using a NIST GE180 sealed cell\footnote{The cell is on loan from T. Gentile at NIST.}, which is filled with 100 torr of $^3$He, 50 torr of N$_2$, and 534 torr of $^4$He, at room temperature. From these studies we extract the magnetic field gradient at 300K 
at the position of the measurement cell. The $^3$He relaxation time at 1.9 K due to this magnetic field gradient is calculated to be $5.26 \times 10^{5}$
seconds. Therefore, the surface effect 
at the walls is the most important contribution to 
the $^3$He relaxation time in our measurements.

The initial improvement observed in the $^3$He relaxation time 
shown in 
Figure~\ref{results} can be attributed to the formation 
of a superfluid $^4$He film on the dTPB-dPS coated acrylic wall. 
However, the thickness of this 
film varies extremely slowly with the amount of $^4$He~\cite{filmref} for the 
entire range of our measurement. 
The behavior of the $^3$He relaxation time as the amount of $^4$He is 
increased, can be understood using simulations described
below~\footnote{The COMSOL Multiphysics finite element package is used
to solve the diffusion equations in our analysis.}.

The $^4$He atoms liquefy and collect at the bottom of the cell with height ranging between 0.0675 mol (0.17cm) and 1.076 mol (2.71cm)
as shown in Figure~\ref{results}. 
The approximate time required for diffusion over a given distance
$h$ is $\frac{h^2}{2 D}$.
The estimated diffusion time of $^3$He from the top of the liquid surface
to the bottom ranges between~$\sim$~60 and 15300 seconds as the height of
the liquid increases.  Our system was far from equilibrium
for most of the measurements we took. A time-dependent diffusion model and a
static magnetic field model are developed to simulate
the signal in the pickup coil in our experiment.

The inset in Figure~\ref{results} shows the measured $^3$He relaxation time
versus the amount
of $^3$He (in mols) in the cell for a 
fixed amount of (0.404 moles) $^4$He. Our results show that the relaxation 
time ($\sim$1800s) is almost independent of the amount of $^3$He in the range 
of our measurement (0.00056 mol to 0.0086 mol).
The model used to analyze the data assumes that 
all $^3$He atoms are in the vapor state 
immediately after the $^3$He
atoms enter the acrylic cell. The concentration is assumed to
be uniform in the vapor, and zero in liquid $^4$He.
$^3$He atoms diffuse both in the vapor and liquid, in which the diffusion
coefficients are different.
At 1.9K, $D_l =2.4 \times 10^{-4}$ cm$^2$/s~\cite{diffref} is the diffusion coefficient of
$^3$He in liquid $^4$He.
The vapor $^3$He diffusion coefficient is calculated using 
$D_v = 1.463 \times 10^{-3} T^{1.65} P^{-1} = 0.018 $ cm$^2$/s~\cite{nacher}.

The boundary condition at the liquid surface is
written using the flux exchange between the vapor and liquid.
The flux going from vapor to liquid is $|\vec{j}_{vl}|=\frac{1}{4} v_v n_v$
and in the opposite direction $|\vec{j}_{lv}|=\frac{v_v}{4}(\frac{m}{m^*})^{3/2}e^{-\frac{E_B}{k T}} n_l$.
$n_v$ and $n_l$ are the concentration of the polarized $^3$He atoms in the vapor
and liquid, respectively. The average speed of $^3$He in the vapor
is $v_v = \sqrt{\frac{8 k T}{\pi m_3}}=1.15\times 10^4$cm/s.
The effective mass of $^3$He dissolved in superfliud $^4$He is $m^{*}_3=2.4m_3$
, where $m_3$ is the mass of a $^3$He atom.

The pickup coil is mounted at the bottom of the acrylic cell and it measures the change of the magnetic flux caused by the spin-flip of the $^3$He magnetic dipoles in the cell (both in the vapor and in the liquid) during an NMR-AFP sweep. In order to calculate this flux, we use the reciprocity theorem, according to which the flux through the pickup coil can be calculated as proportional to the field produced by a current in the pickup coil at the location of the $^3$He dipole.

\begin{figure*}
\begin{center}
\includegraphics[width=18cm]{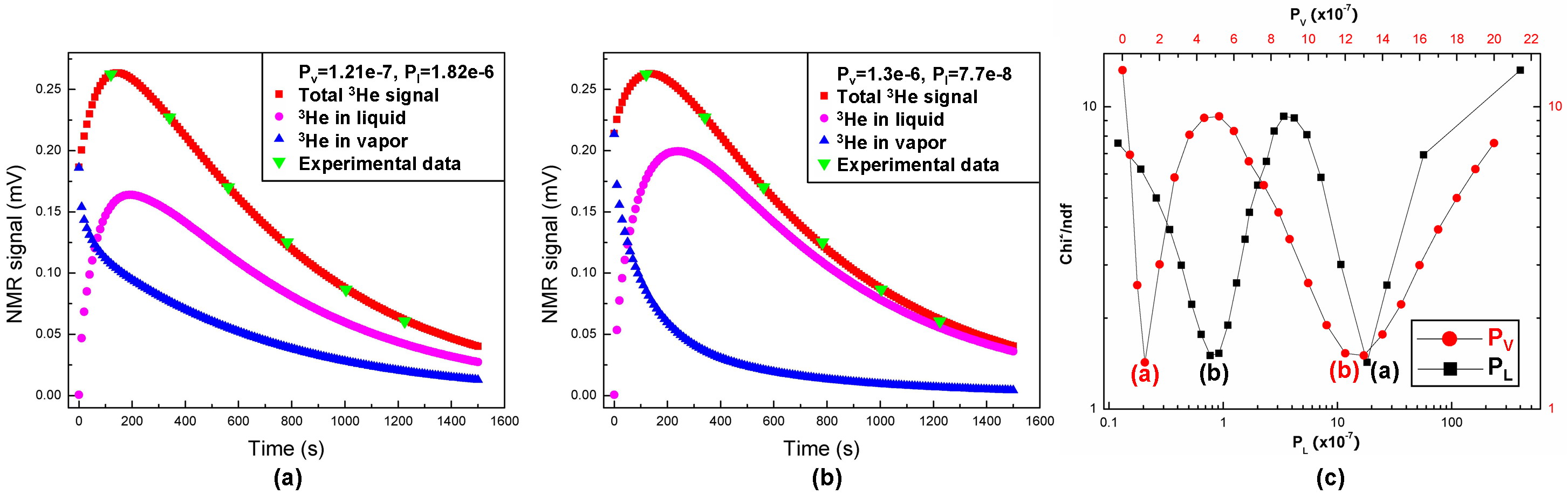}
\caption{(a) and (b) are NMR measurements of the $^3$He signal (green triangles) at 1.9K as a function of time with the amount of $^4$He equal to 0.135 mol (0.34 cm). Red squares are the simulated total signal in the pickup coil consisting of the contributions from the vapor (blue triangles) and liquid (pink circles). (c) is reduced $\chi^2$ obtained from the best fit as a function of $P_v$ (red circles, top axis) and $P_l$ (black squares, bottom axis) showing how different values of $P_v$ and $P_l$ can fit the data 
due to the fact that with low liquid level the vapor is close to the pickup coil.}
\label{fig:034run}
\end{center}
\end{figure*}

\begin{figure}
\begin{center}
\includegraphics[width=7.5cm]{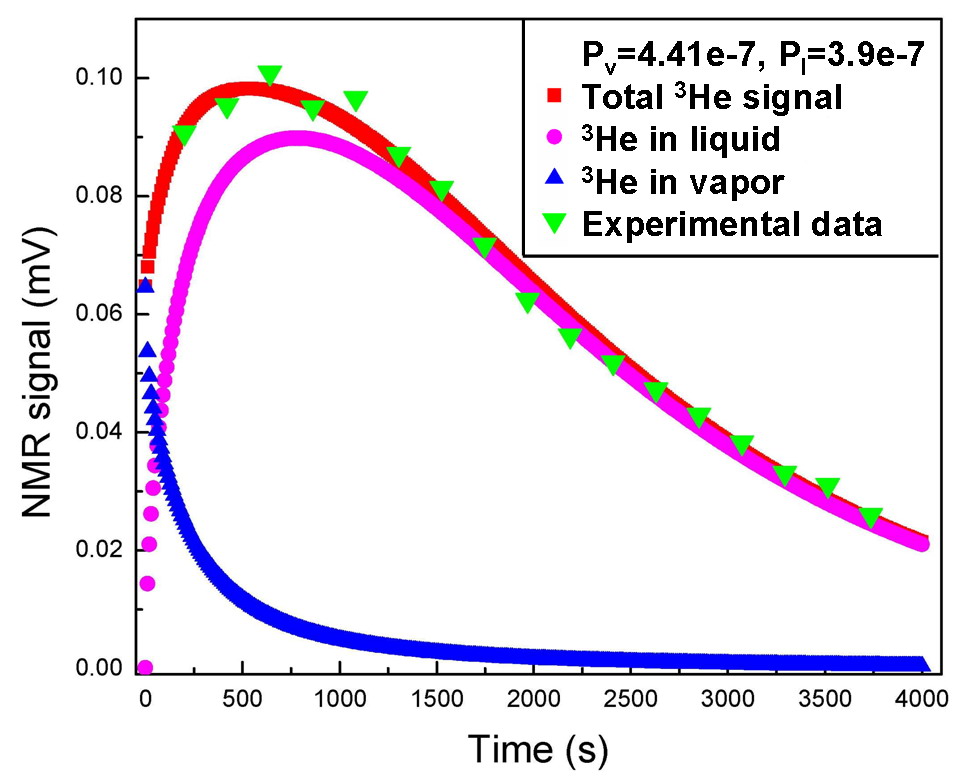}
\caption{The amount of $^4$He in the acrylic cell is 0.673 mol (1.69 cm). The experimental data (green triangles) consisting of the contributions from the vapor (blue triangles) and liquid (pink circles) are fitted onto the simulation results (red squares).}
\label{fig:169run}
\end{center}
\end{figure}

\begin{figure}
\begin{center}
\includegraphics[width=7.5cm]{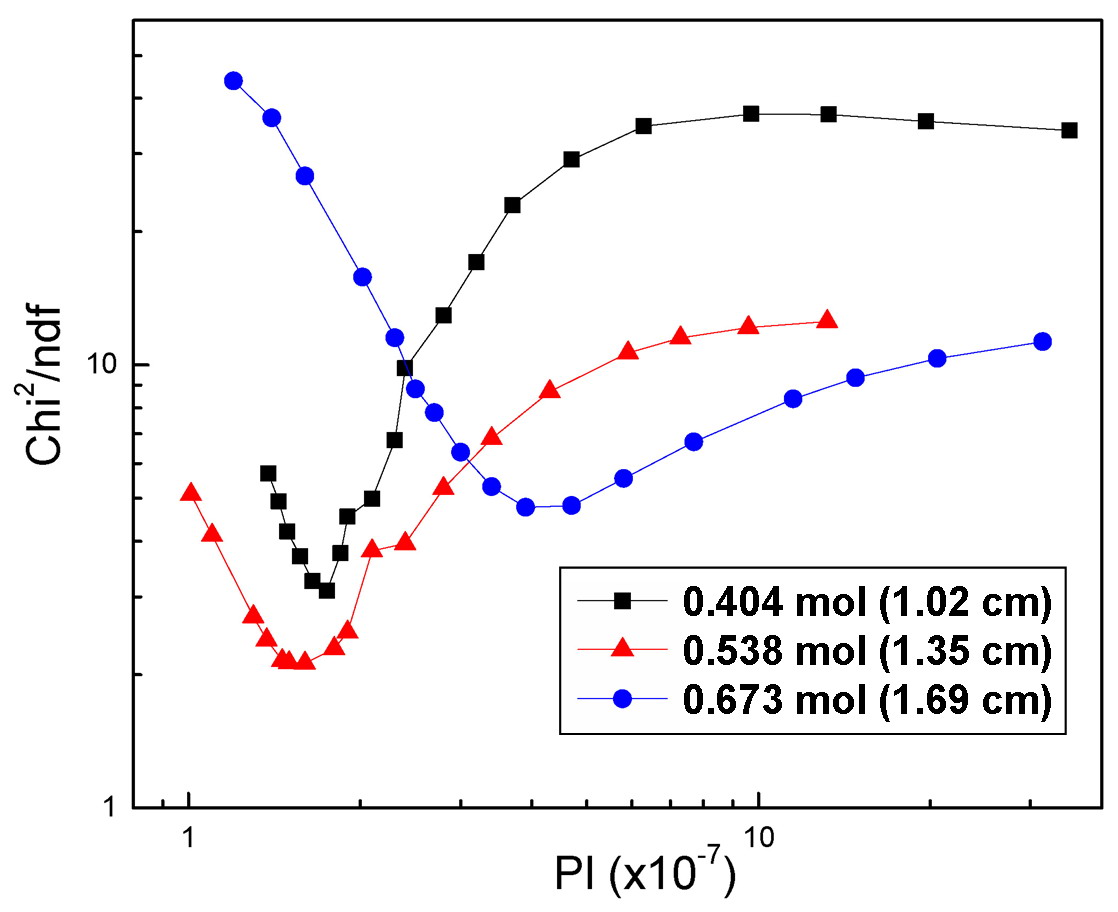}
\caption{Reduced $\chi^2$ for the fits versus $P_l$ values for $^4$He amounts of 0.404 mol (1.02 cm), 0.538 mol (1.35 cm) and 0.673 mol (1.69 cm).}
\label{fig:chi2ndf}
\end{center}
\end{figure}

Our measurements are best characterized by the depolarization probability (DP) per wall collision. In the analysis we allow for this probability to be different on the walls covered with bulk liquid, $P_{l}$, and the walls covered with superfluid film only, $P_{v}$.
The wall boundary condition is that depolarization rate on the wall is the product of the number of atoms reaching the wall per unit time and the corresponding DP. The depolarization behavior of the model can be changed by varying these parameters.
To extract these parameters we did least squares fits of the measured $^3$He NMR-AFP signal versus time from injection of the polarized $^3$He into the $^4$He containing cell, with the same quantity calculated by solving the diffusion equation as specified above and using the solutions to calculate the signal in the pickup coil. 

Because $^3$He atoms dissolve into the liquid $^4$He rapidly without losing polarization, the signal increases from zero in the beginning of the measurement and then decays after it saturates.
$P_{v}$ will influence the short time buildup of the signal in the pickup coil and $P_{l}$ will determine the long time behavior. Thus a larger $P_{v}$ and smaller $P_{l}$, will move the peak of the signal to shorter times.
The pickup coil is located on the bottom of the measurement cell and so is more sensitive to $^3$He dissolved in the liquid. And it becomes less sensitive to the $^3$He in the vapor as the amount of $^4$He is increased.  In the measurements made with small amounts of $^4$He the long time behavior is influenced by both $P_{v}$ and $P_{l}$. So it is difficult to extract unique values of the parameters from these data. As an example Figure~\ref{fig:034run} (a) and (b) show the varying contribution of the $^3$He in the liquid (pink circles) and vapor (blue triangles) to two equally good fits (red squares and green triangles) for $P_{v,l}$ varying by about a factor of 10. Figure~\ref{fig:034run} (c) shows the plot of reduced $\chi^{2}$ obtained from the best fits as a function of $P_{l}$ (bottom axis) and $P_{v}$ (top axis).
For larger $^4$He quantities the results are not sensitive enough to $P_{v}$ to allow the extraction of a value for this quantity. 

The fitting is made more difficult in that we have no absolute polarization information so that we have to treat the normalization of the curves as a free parameter. This is another reason that a range of parameters can give good fits in the low filling cases. In addition, our operational procedures were such that in most cases we started taking data after the peak had been passed. Only the run with 0.673 mol (1.69 cm) of $^4$He shows the peak of the signal (Figure~\ref{fig:169run}) and we are able to extract a reasonable value of $P_{l}$ from the fit (Figure~\ref{fig:chi2ndf}), $P_{l}=(3.9^{+2.0}_{-0.7}) \times 10^{-7}$.	
The error bar is determined by the standard method of varying the $P_{l}$ parameter so that the reduced $\chi^2$ is increased by 1.
For measurements with larger amounts of $^4$He, diffusion to the walls plays a significant role and calculations show that the long time
behavior is less sensitive to the value of $P_{l}$ so we cannot extract meaningful values of the wall loss parameters from the data.
In Figure~\ref{fig:chi2ndf}, we also show the reduced $\chi^{2}$ plots for the runs with 0.404 mols (1.02 cm) and 0.538 mols (1.35 cm). From these plots we can extract $P_{l}=(1.7 \pm 0.2) \times 10^{-7}$ and $P_{l}= (1.6 \pm 0.4) \times 10^{-7}$ respectively. The minima in reduced $\chi^{2}$ are much broader when plotted versus $P_{v}$ so we cannot extract useful values for this parameter.

Lusher {\it et al.}~\cite{lusher} carried out a series of measurements with open Pyrex glass chambers as well as sealed Pyrex glass cells. Their results showed that the formation of a superfluid $^4$He film on a hydrogen coated glass surface reduces the depolarization of $^3$He from the surface. For an open cell they observed a relaxation time of $\sim$500 seconds at a magnetic holding field of 0.23 Tesla and a temperature of 1.9 K. The $^3$He bulk number density for these measurements was $5.2 \times 10^{-6}$ mol/cc (cell volume 4.2 cc) and the $^3$He~:~$^4$He atomic ratio was 1:16 (ours is 1:769). As shown in Figure~\ref{results} we have observed relaxation times in excess of 3000 seconds at 1.9~K for a holding field of 21 Gauss. The surface to volume ratio of our cell is 50\% of the cells used in measurements of ~\cite{lusher}, and our measured relaxation time is a convolution of $^3$He $T_1$ and the $^3$He diffusion time constant. 
Their corresponding depolarization probability is determined to be $\sim 1.9 \times 10^{-7}$, which is similar to our $P_l$ value, though ours is obtained
from a dTPB-dPS coated acrylic surface under the superfluid $^4$He liquid.

We have measured the relaxation time
of polarized $^3$He in a dTPB-dPS coated acrylic cell in a 
diluted mixture of $^3$He-$^4$He at 
a temperature of 1.9 K 
with a magnetic holding field of 21 Gauss.
We have shown that it's possible to achieve values of wall depolarization probability ($P_{l}$) on the order of $(1 - 2)\times 10^{-7}$ for polarized $^3$He in the superfluid $^4$He at 1.9K.
To provide precise 
determination of these depolarization probabilities in future measurements, 
one needs to isolate the diffusion time scale from the system, i.e. to 
carry out measurements in a cell with superfluid $^4$He film on the wall only, 
and measurements from a cell filled with superfluid $^4$He completely. 
It also remains to be seen how sensitive depolarization probabilities are to surface preparations.
Nevertheless, ours is the first study of the 
polarized $^3$He relaxation time
from dTPB-dPS coated surfaces in superfluid $^4$He.
Our data suggest that such surface 
may find applications 
in areas which employ polarized $^3$He at low temperatures
in the environment of superfluid $^4$He.
Since the $^3$He behavior is mostly dominated by diffusion in liquid $^4$He at 1.9K, it is important to extend our current work to below 1K due to much shorter $^3$He diffusion time. Such measurements are currently in progress.

We thank R.P. Behringer, T. Clegg, D. Haase, G. Finkelstein,
T. Gentile, M. Hayden, V. Cianciolo, P. Huffman, T.~Katabuchi, H. Meyer, and A.~Tobais for helpful discussions. 
We also thank J.~Rishel for making all the glassware 
for these measurements and 
N. Boccabello, B.~Carlin, J. Dunham, R. O'Quinn, P.~Mulkey and C. Westerfeld for the technical support.
This work is supported in part by the School of Arts and Science of 
the Duke University, the U.S. Department of Energy under 
contract number DE-FG02-03ER41231, and additional support from DOE.


\end{document}